\documentclass[aps,prl,twocolumn,superscriptaddress,notitlepage,nofootinbib,longbibliography]{revtex4-2}
 \usepackage{mathrsfs}
 \usepackage{epsfig}
 \usepackage{graphicx}
 \usepackage{amsfonts}
 \usepackage[figuresright]{rotating}
 \usepackage{amssymb}
 \usepackage{amsmath}
 \usepackage{dcolumn}
 \usepackage{bm}
 \usepackage{xcolor}
 \usepackage{color}
 \usepackage{braket}
 \usepackage{units}
 \usepackage{xspace}

 \usepackage{bbold}
 \definecolor{mypink3}{cmyk}{0, 0.7808, 0.4429, 0.1412}
 \definecolor{mypink1}{rgb}{0.858, 0.188, 0.478}
 \definecolor{mypink2}{RGB}{219, 48, 122}
 \usepackage[colorlinks=true, allcolors=mypink1]{hyperref}
 \usepackage[fleqn]{mathtools}

 \usepackage[shortlabels]{enumitem}
 \newcommand{\ECNU}{Quantum Institute for Light and Atoms, School of Physics, East China Normal University, Shanghai 200062, China}
 \newcommand{\SBH}{Shanghai Branch, Hefei National Laboratory, Shanghai 201315, China}
 \newcommand{\SPA}{School of Physics and Astronomy, and Tsung-Dao Lee Institute, Shanghai Jiao Tong University, Shanghai 200240, China}
 \newcommand{\SRC}{Shanghai Research Center for Quantum Sciences, Shanghai 201315, China}
 \newcommand{\CIC}{Collaborative Innovation Center of Extreme Optics, Shanxi University, Taiyuan, Shanxi 030006, China}

\begin{document}

\title{Spatiotemporal Non-Hermitian Skin Effect with Floquet-Engineered Ultracold Atoms}
	
 \author{Weijie Liang}
 \affiliation{\ECNU}
 \author{Weiping Zhang}
 \email{wpz@sjtu.edu.cn}
 \affiliation{\SPA}
 \affiliation{\SBH}
 \affiliation{\SRC}
 \affiliation{\CIC}
 \author{Keye Zhang}
 \email{kyzhang@phy.ecnu.edu.cn}
 \affiliation{\ECNU}
 \affiliation{\SBH}

\begin{abstract}

Non-Hermitian skin effect, the accumulation of bulk eigenstates at edges under open boundary conditions, has been observed across classical, quantum, and synthetic platforms. Whether it can be generalized to the genuine time dimension remains open. We propose a spatiotemporal NHSE in the Floquet phase space via Floquet engineering of a ring-trapped ultracold atomic gas. Placing spatial and temporal degrees of freedom on an equal footing, we show that a single parent skin effect projects onto either domain. The spatial projection yields an NHSE with arbitrarily tunable skin locations, eliminating the need for physical open boundaries. The temporal projection gives time-domain nonreciprocity and periodic temporal funneling, consistent with causality. This establishes a unified spatiotemporal paradigm for NHSE and opens a route to non-Hermitian topological physics in time.

\end{abstract}
\maketitle

\emph{Introduction.}---The non-Hermitian skin effect (NHSE), originally formulated within non-Hermitian quantum mechanics~\cite{PhysRevLett.116.133903,PhysRevB.97.121401,PhysRevLett.121.086803}, has since been extensively observed across diverse classical wave platforms that emulate wavefunctions, including photonic lattices~\cite{weidemann2020topological,wang2021generating,PhysRevApplied.14.064076}, acoustic cavities~\cite{zhang2021observation,zhang2021acoustic,gu2022transient}, and electrical circuits~\cite{helbig2020generalized,liu2021non,PhysRevB.107.085426,PhysRevResearch.4.033109}. More recently, its study has expanded to genuine quantum platforms, such as atomic Bose--Einstein condensates~\cite{PhysRevLett.124.070402,liang2022dynamic} and superconducting quantum circuits~\cite{shen2025observation}. The NHSE exhibits singular features absent in its Hermitian counterparts, most notably topologically protected and self-healing skin modes~\cite{PhysRevLett.128.157601} and directional funneling toward boundaries~\cite{weidemann2020topological,PhysRevLett.125.126402,lin2022observation,weidemann2022topological}. These unconventional dynamics have unlocked versatile applications ranging from robust light transport and energy harvesting~\cite{weidemann2020topological,longhi2015robust,longhi2020stochastic,lin2023manipulating} to programmable morphing of topological modes~\cite{wang2022non} and fault-tolerant quantum information processing~\cite{tang2025symmetry}. 

Beyond physical space, the NHSE has also been extended to synthetic dimensions, including frequency and momentum networks~\cite{song2020two, liang2022dynamic, PhysRevLett.129.243901}. With recent breakthroughs in time crystals~\cite{PhysRevLett.109.160401,PhysRevLett.117.090402,zhang2017observation,yao2020classical,kongkhambut2022observation}, temporal moiré lattices~\cite{thvw-pdtd}, and time-domain topological phases~\cite{lustig2018topological,ren2025observation,xiong2025observation,feis2025space,tong2025observation,9b46-d2ry}, it is natural to ask whether the NHSE can be generalized to the genuine time dimension, and what distinct physics a time-domain skin effect would exhibit compared to its spatial counterpart. Such an extension would allow the conceptual tools developed for spatial non-Hermitian quantum control to be applied in the temporal domain.

However, extending the NHSE into the genuine temporal dimension faces a fundamental conceptual challenge. The conventional spatial NHSE relies on two essential ingredients, namely nonreciprocal couplings and physical open boundary condition (OBC)~\cite{PhysRevLett.127.116801,zhang2022universal,zhao2025two}. In the time domain, causality precludes any backward coupling channel, rendering genuine nonreciprocal temporal driving highly nontrivial. Moreover, unlike a spatial boundary, the arrow of time cannot be halted or reflected. These constraints appear to preclude a straightforward temporal counterpart of the NHSE. Recent studies have attempted to circumvent them by creating temporal nonreciprocity through velocity-dependent dissipation~\cite{gfy2-d4pp} or by simulating skin-like dynamics within artificial lattices mimicking the time dimension~\cite{PhysRevLett.134.243805}. Nevertheless, achieving a genuine temporal skin effect in a real, continuous temporal degree of freedom, while clearly resolving its similarities and differences with its spatial counterpart, remains an open challenge.

In this work, we overcome these challenges by introducing the concept of spatiotemporal NHSE, together with a realistic implementation scheme in a periodically driven, ring-trapped ultracold atomic system. By elevating the NHSE framework into the Floquet phase space (FPS) equipped with a co-moving spatiotemporal frame~\cite{buchleitner2002non}, spatial and temporal degrees of freedom are placed on an equal footing. This allows a single parent skin effect to be projected onto either domain, thereby selectively realizing both spatial and temporal NHSE. Crucially, the spatial projection yields an NHSE with arbitrarily tunable skin locations under periodic boundary conditions (PBC), without requiring spatial open boundaries. In parallel, the temporal projection manifests as time-domain nonreciprocity and a periodic temporal funneling effect, both strictly consistent with the causality and the boundaryless nature of time. Our work thus establishes a unifying spatiotemporal paradigm for non-Hermitian skin dynamics, opening unexplored avenues for non-Hermitian topological physics in the genuine temporal dimension and for spatiotemporal quantum control.

\begin{figure}[t]
	\centering
	\includegraphics[width=0.8\linewidth, height=0.82\linewidth]{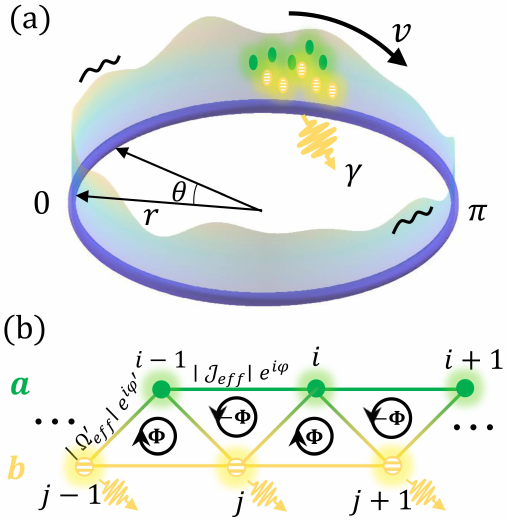}
	\caption{(a) Ultracold atoms rotating at velocity $v$ in a ring trap of radius $r$, subjected to a time-periodic spatial perturbation (shaded cylinder with a modulated top edge). Two internal atomic states are coupled, with state $\vert{}\psi_b\rangle$ (striped circle) subject to dissipation at rate $\gamma$. (b) Zigzag ladder synthesized by the driving protocol of (a) in the Floquet phase space. Clockwise and counterclockwise circulations around each triangular plaquette acquire opposite geometric phases $\pm\Phi$ (circled arrows). The upper and lower legs denote the internal states $|\psi_a\rangle$ and $|\psi_b\rangle$, respectively.}
	\label{fig1}
\end{figure}

\emph{Model.}---We consider two-state ultracold atoms in a ring trap of radius $r$, rotating at average velocity $v$ [Fig.~\ref{fig1}(a)]. The ring geometry imposes spatial PBCs, while the rotation introduces temporal periodicity. Their interplay yields a unified boundary condition in spatiotemporal coordinates, which we introduce later. The atoms undergo static inter-state coupling and intra-state Floquet engineering via temporal periodic perturbations. In the rotating frame, the Hamiltonian along the angular coordinate $\theta$ in the basis $(\vert{}\psi_a\rangle, \vert{}\psi_b\rangle)^{\mathrm{T}}$ reads
\begin{equation}
\label{Eq1}
\mathcal{H}_0=
\begin{pmatrix}
\frac{\hat p^2}{2Mr^2}+\mathcal{V}_a(\theta,t) & \hbar\Omega_R \\
\hbar\Omega_R & \frac{\hat p^2}{2Mr^2}+\mathcal{V}_b(\theta,t)-i\hbar\gamma
\end{pmatrix},
\end{equation}
where $\hat p$ is the angular momentum, $\Omega_R$ is the coupling Rabi frequency, and $\gamma$ is the dissipation rate of state $|\psi_b\rangle$. Interatomic collisions are neglected to rule out nonlinear localization, and the two internal states are assumed degenerate for simplicity.

The time-periodic perturbations are given by $\mathcal{V}_{a,b}(\theta,t)=V(\theta)F_{a,b}(t)$, comprising an arbitrary PBC-preserving spatial potential $V(\theta)$ and state-dependent periodic driving functions $F_{a,b}(t)$. Their double Fourier expansion reads $\mathcal{V}_{a,b}=\sum_{n,m}V_n F_{m} e^{i(n\theta-m\omega t)}$. Substituting the unperturbed angular motion $\theta=\Omega t$ (associated with the unperturbed angular momentum $p_0=Mrv$ via $\Omega=v/r$) yields phase factors $e^{i(n\Omega-m\omega)t}$. Consequently, the dynamics are dominated by near-resonant components satisfying $n\Omega \approx m\omega$, where slow phase variations enable coherent coupling while off-resonant terms rapidly average out. To isolate the effective dynamics, we apply a secular approximation that eliminates these off-resonant contributions and transform to the FPS defined by the slow variables $\hat P=\hat p-p_0$ and $\bar{\theta}=\theta-\Omega t$. This allows time to act as a synthetic spatial dimension, a framework commonly employed in atomic time-crystal studies \cite{PhysRevLett.111.205303,PhysRevA.91.033617,thvw-pdtd}. 
As we show below, properly designed driving functions $F_{a,b}(t)$ generates in FPS a lattice structure, an artificial open boundary, and a non-reciprocal hopping. None of these ingredients exist in the laboratory frame. Their combination produces an engineered spatiotemporal NHSE.

We consider a composite Floquet driving protocol $F_{a,b}(t)=\cos[\omega t+\phi(t)+\Theta_{a,b}]+f_e(t)$ to engineer the spatiotemporal lattice and boundary in the FPS. The frequency-multiplied cosine term ($\omega=s\Omega$, $s\in\mathbb{Z}^+$), combined with the resonant Fourier component of the spatial potential $V(\theta)$, generates an effective $s$-site lattice potential $V_s \cos[s\bar{\theta}-\phi(t)-\Theta_{a,b}]$ in the FPS, where the slowly varying phase $\phi(t)$ enables dynamical modulation and the phase offsets $\Theta_{a,b}$ introduce a relative lattice shift between the two internal states. Simultaneously, the second term, with Fourier expansion $f_e(t)=\sum_n F_n e^{in(\Omega t+\Theta_e)}$, can construct generalized boundary conditions for an arbitrary $V(\theta)$~\cite{PhysRevLett.127.116801}. Specifically, choosing $V_n F_n = \exp(-n^2/2a^2)$ yields a Gaussian barrier $V(\theta)f_e(t)\approx V_e \exp[-a^2(\bar{\theta}-\Theta_e)^2/2]$ that suppresses inter-domain tunneling across $\bar{\theta}=\Theta_e$ in the FPS, approaching an open boundary condition in the strong-barrier regime ($V_e\gg V_s$). Thus, the FPS framework allows flexible lattice and boundary engineering even when the laboratory-frame potential $V(\theta)$ is non-lattice and PBC-preserving (see Supplemental Material for several examples ~\cite{supplementarymaterial}).

At first glance, a homogeneous lattice potential in the FPS appears incompatible with the NHSE. However, setting $\Theta_b - \Theta_a = \pi$ shifts the lattice potential for $\lvert\psi_b\rangle$ by half a period relative to $\lvert\psi_a\rangle$, mapping the system onto a continuum zigzag ladder. To induce non-reciprocal hoppings, we introduce a low-frequency two-tone phase shaking $\phi(t) = A_1\sin(2\omega_s t) + A_2\cos(\omega_s t)$ ($\omega_s \ll \Omega$), whose period $T_s = 2\pi/\omega_s$ is an integer multiple of $T_0 = 2\pi/\Omega$. This drive generates synthetic Peierls phases~\cite{struck2012tunable} that accumulate along the zigzag geometry into a chiral geometric phase. Together with state-selective dissipation $\gamma$, this chiral flux breaks spatial-reflection symmetry to yield non-reciprocal hoppings, opening a complex topological point gap that drives the NHSE under effective boundary confinement.

To make this physics explicit, we transform $\mathcal{H}_0$ to the FPS and apply a unitary transformation that absorbs the time-dependent phase modulation into a dynamical lattice tilt. The resulting Hamiltonian is
\begin{equation}
\mathcal{H}_1 = \frac{\hat P^2}{2Mr^2}\mathbf{I} +
\begin{pmatrix}
\mathcal{V}_a(\bar{\theta})+\mathcal{F}(t)\bar{\theta} & \hbar\Omega_R \\
\hbar\Omega_R & \mathcal{V}_b(\bar{\theta})+\mathcal{F}(t)\bar{\theta}-i\hbar\gamma
\end{pmatrix},
\end{equation}
where $\mathcal{V}_{a,b}(\bar\theta)$ are static lattice potentials shifted by half a period, and $\mathcal{F}(t)=\frac{Mr^2}{s}\frac{d^2\phi(t)}{dt^2}$ represents an effective inertial force induced by the phase shaking.

When the shaking frequency satisfies $4 E_r (V_s/E_r)^{3/4} e^{-2\sqrt{V_s/E_r}} / \sqrt{\pi} +\hbar\Omega_R \ll \hbar\omega_s \ll \sqrt{2 V_s E_r}$, with recoil energy $E_r=\hbar^2 s^2 / (2 M r^2)$, higher bands are isolated, which validats a tight-binding description restricted to the lowest two bands. Expanding the atomic field operators in the Floquet-Wannier basis yields the effective lattice Hamiltonian
\begin{equation}
\label{H2}
\begin{aligned}
    &\mathcal{H}_2=-|\mathcal{J}_{\text{eff}}|\sum_{i,j}( e^{i\varphi}a_i^\dagger a_{i+1}+e^{i\varphi}b_j^\dagger b_{j+1}+h.c.)-i\hbar\gamma\sum_j n_{j}^b\\
    &+\sum_{\langle i,j\rangle}\hbar|\Omega_{\text{eff}}^\prime|(e^{-i\varphi^{\prime}}a_i^\dagger b_j+e^{i\varphi^{\prime}}a_i^\dagger b_{j+1}+h.c.)
    +\mathcal{H}_e,
\end{aligned}
\end{equation}
where $\vert\mathcal{J}_{\text{eff}}\vert e^{i\varphi}$ and $\vert{}\Omega^\prime_{\text{eff}}\vert{}e^{i\varphi'}$ denote the effective intra-leg hopping and inter-leg coupling, respectively, with synthetic Peierls phases $\varphi$ and $\varphi'$ controlled by driving parameters $A_{1,2}$ and $\omega_s$ (see Supplemental Material for derivations~\cite{supplementarymaterial}). Here, $a_i$ ($b_j$) annihilates an atom in state $\vert{}\psi_a\rangle$ ($\vert{}\psi_b\rangle$) at site $i$ ($j$), $n_i^{a,b}$ are occupation operators, and $\mathcal{H}_e$ characterizes the engineered generalized boundary generated by the potential barrier $V_e$.

In the zigzag ladder, the effective hopping between adjacent sites $a_i$ and $a_{i+1}$ arises from the interference between the direct path $a_i \to a_{i+1}$ and the two-step path $a_i \to b_j \to a_{i+1}$ via the lower-leg site $b_j$ [Fig.~\ref{fig1}(b)]. These two paths enclose a triangular plaquette with a relative geometric phase that is strictly chiral, accumulating $+\Phi = 2\varphi' - \varphi$ for rightward propagation ($a_i \to a_{i+1}$) and $-\Phi$ for leftward propagation ($a_{i+1} \to a_i$). In the absence of dissipation ($\gamma = 0$), this sign reversal merely induces direction-opposite phase shifts without altering the hopping magnitudes. However, once the dissipation $\gamma$ on state $\vert{}\psi_b\rangle$ is introduced, non-Hermitian path interference causes the chiral phase $\pm\Phi$ to interfere constructively in one direction and destructively in the other. This functions as a direction-dependent filter that converts phase chirality into amplitude asymmetry, directly generating non-reciprocal effective hoppings ($\vert{}t_R\vert{} \neq \vert{}t_L\vert{}$) along the ring. A detailed analytical derivation of these asymmetric hopping amplitudes via the adiabatic elimination of state $\vert{}\psi_b\rangle$ is provided in the Supplemental Material~\cite{supplementarymaterial}.

\begin{figure}[t]
	\centering
	\includegraphics[width=1\linewidth, height=0.78\linewidth]{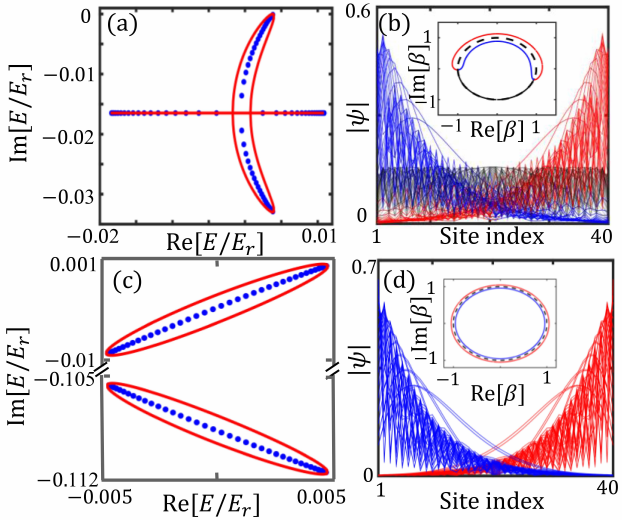}
	\caption{Complex energy spectra under PBC (red solid) and OBC (blue dotted) for (a) $\gamma=3\vert{}\Omega_R'\vert{}$ and (c) $\gamma=10\vert{}\Omega_R'\vert{}$. (b),(d) Spatial profiles of OBC eigenstates $\vert{}\psi_{a,b}\vert{}^2$ corresponding to (a) and (c), respectively. Insets: GBZ in the complex $\beta$-plane (blue/red solid) and the unit circle (black dashed). Parameters: $\vert{}\mathcal{J}_{\text{eff}}\vert{}=0.002E_r$, $\hbar\vert{}\Omega_{\text{eff}}'\vert{}=5.5\vert{}\mathcal{J}_{\text{eff}}\vert{}$, $\varphi=1.47$, $\varphi'=-2.25$, $s=40$, and $E_r=\hbar^2 s^2 / (2 M r^2)$.
    }
	\label{fig3}
\end{figure}

In the absence of the boundary term $\mathcal{H}_e$, the system preserves translation invariance under PBC, and $\mathcal{H}_2$ can be diagonalized in quasi-momentum $k$-space, yielding the complex dispersion, $E_\pm(k) = -2\vert{}\mathcal{J}_{\text{eff}}\vert{}\cos(kd+\varphi) - i\frac{\hbar\gamma}{2} \pm \hbar\sqrt{4\vert{}\Omega_{\text{eff}}'\vert{}^2\cos^2\left(\frac{kd}{2}+\varphi'\right) - \frac{\gamma^2}{4}}$ with $d=2\pi/s$ being the lattice constant. 
Note that the constant imaginary term $-i\hbar\gamma/2$ merely induces an overall energy shift without affecting the topological properties of the complex energy spectrum. 

When both the dissipation rate $\gamma \neq 0$ and the net geometric phase $\Phi = 2\varphi' - \varphi \neq 0$, the two energy branches $E_\pm(k)$ trace out closed loops in the complex energy plane as $k$ spans the Brillouin zone. As shown in Figs. \ref{fig3}(a) and \ref{fig3}(c), these loops enclose finite spectral regions, forming non-trivial topological point gaps (red solid lines)~\cite{PhysRevX.8.031079,PhysRevX.9.041015}. Since any system under OBC is strictly point-gap trivial, this topological mismatch forces the spectral loops to collapse inward into open curves (blue dotted lines) when $\mathcal{H}_e$ is present, inevitably triggering the NHSE \cite{PhysRevLett.124.086801}. 

The square root in $E_\pm(k)$ introduces an exceptional threshold at $4\vert{}\Omega_R'\vert{} = \gamma$. Above this threshold ($4\vert{}\Omega_{\text{eff}}'\vert{} > \gamma$), the PBC spectrum consists of coexisting open curves and closed loops [Fig. \ref{fig3}(a)], where the corresponding OBC eigenstates display a coexistence of extended states (black lines) and boundary-localized skin modes (blue and red lines) [Fig. \ref{fig3}(b)]. 
Below it ($4\vert{}\Omega_{\text{eff}}'\vert{} < \gamma$), the entire PBC spectrum forms closed loops [Fig. \ref{fig3}(c)], rendering all OBC eigenstates exponentially localized at the boundaries [Fig. \ref{fig3}(d)]. 

This localization behavior is further corroborated by non-Bloch band theory, through calculating the generalized Brillouin zone (GBZ) for the complex wavevector $\beta = e^{ik_\beta}$ under OBC \cite{PhysRevLett.121.086803, PhysRevLett.123.066404}. When $4\vert{}\Omega_{\text{eff}}'\vert{} > \gamma$, the GBZ lies partly inside ($\vert{}\beta\vert{} < 1$, blue line) and partly outside ($\vert{}\beta\vert{} > 1$, red line) the unit circle while partially coinciding with it ($\vert{}\beta\vert{} = 1$, black line) [inset of Fig. \ref{fig3}(b)], signifying left- and right-localized skin modes coexisting with extended states. In contrast, when $4\vert{}\Omega_{\text{eff}}'\vert{} < \gamma$, the GBZ strictly deviates from the unit circle everywhere [inset of Fig. \ref{fig3}(d)], confirming that all eigenstates undergo skin localization.

When $\varphi = 2\varphi'$ (i.e., vanishing geometric phase $\Phi = 0$), the phase alignment between the cosine terms locks the real and imaginary parts of $E_\pm(k)$ along a one-dimensional trajectory. Consequently, the PBC spectrum no longer exhibits a topological point gap, suppressing the NHSE even in the presence of boundaries.
We note that analogous topological non-reciprocity can be realized by engineering atomic gain/pumping instead of dissipation $\gamma$, however, precise experimental control over atomic pumping rates remains challenging.

\begin{figure}[t]
	\centering
	\includegraphics[width=0.95\linewidth, height=1.14\linewidth]{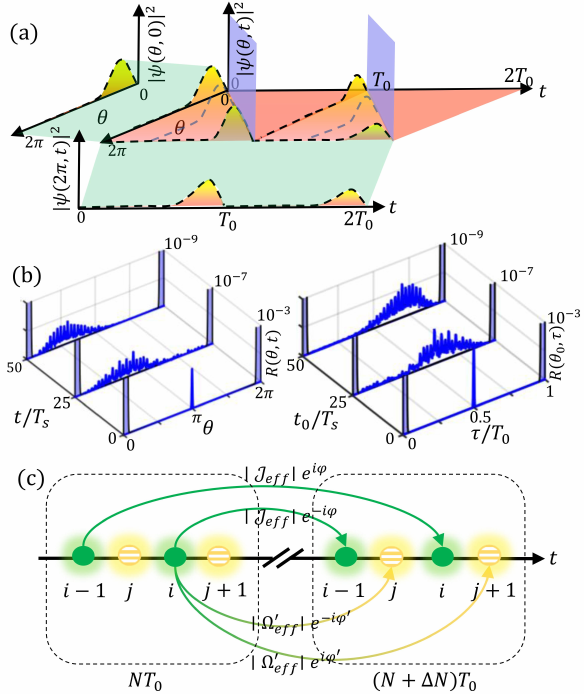}
	\caption{(a) Spatiotemporal NHSE dynamics with skin-mode wave functions mapped onto temporal and spatial domains. (b) Probability decay rate distributions. Left: spatial distribution $R(\theta,t)$ versus $\theta\in[0,2\pi)$ at stroboscopic times $t= 0, 25 T_s$, amd $50 T_s$. Right: intra-period distribution $R(\theta_0,t_0+\tau)$ with $\tau\in[0,T_0)$ at a fixed position $\theta_0=0$, where $t_0$ takes the same stroboscopic times. Blue shaded regions denote the effective boundaries in both panels. Parameters are: $\Omega = 200 E_r/\hbar$, $\omega = s\Omega$, $\Omega_R = 0.2 E_r/\hbar$, $\omega_s = 0.5 E_r/\hbar$, $A_1 = 0.95$, $A_2 = 1.5$, $V_s = E_r$, $\gamma = 0.05 E_r/\hbar$, with recoil energy $E_r = \hbar^2 s^2 / (2Mr^2)$ and $s = 40$. (c) Schematic of causal temporal hopping, where the forward arrow of time strictly governs both leftward and rightward coupling channels.}
	\label{fig4}
\end{figure}

\emph{Spatiotemporal NHSE.}---In the FPS, the NHSE drives probability accumulation near $\bar{\theta} = \Theta_e$, which maps in the laboratory coordinates $(\theta, t)$ to a family of parallel spatiotemporal lines $\theta - \Omega t = \Theta_e \pmod{2\pi}$ due to the spatial periodicity $\theta \in [0, 2\pi)$ and stroboscopic temporal repetition $t \to t + N T_0$ (where $T_0 = 2\pi/\Omega$) [Fig.~\ref{fig4}(a)]. The mapping preserves chirality in space, $\partial\theta/\partial\bar{\theta}=+1$, but reverses it in time, $\partial t/\partial\bar{\theta}=-1/\Omega$. Its spatial and temporal projections thus yield controllable, chirality-reversed non-Hermitian funneling dynamics without requiring physical open boundary conditions, enabling versatile spatial and temporal skin control~\cite{weidemann2020topological}.

To verify the spatial projection, we simulate the non-Hermitian atomic dynamics in the laboratory frame, starting from a Gaussian wavepacket in $\vert\psi_a\rangle$ centered at $\theta=\pi$ and rotating with an angular frequency $\Omega$ ($\vert\psi_b\rangle = 0$). Due to the ring geometry, when observed stroboscopically at $t=NT_0$, the rising and falling edges of the barrier positioned at $\Theta_e = 0$ act as effective left and right spatial boundaries near $\theta \approx 0$ and $\theta \approx 2\pi$, respectively. Although dissipation $\gamma$ is spatially uniform, non-reciprocal coupling drives atoms to continuously escape the trap at an asymmetric spatial rate $R(\theta,t) \equiv -\frac{d}{dt}\langle \psi(t)\vert{}\psi(t) \rangle = 2\gamma \vert{}\psi_b(\theta,t)\vert{}^2$~\cite{PhysRevLett.128.120401}. As depicted in the left panel of Fig. \ref{fig4}(b) at snapshots $t = 0, 25T_s,$ and $50T_s$, the high-decay regions systematically drift toward and condense at the left boundary, confirming spatial funneling.

Conversely, probing at a fixed spatial position $\theta_0$ translates the FPS skin mode into time-domain funneling. To track its continuous evolution, we slice $\psi(\theta_0, t)$ into stroboscopic periods starting at $t_0 = N T_0$, defining the intra-period temporal wavefunction $\psi_{a,b}(\theta_0, t_0 + \tau)$ with $\tau \in [0, T_0)$. This yields a period-resolved decay rate $R(\theta_0, t_0 + \tau) = 2\gamma \vert{}\psi_b(\theta_0, t_0 + \tau)\vert{}^2$. Analogous to the spatial case evaluated at different evolution times $t$, observing $R(\theta_0, t_0 + \tau)$ for increasing stroboscopic origins $t_0$ captures the dynamical evolution of temporal funneling along the macroscopic time axis. As shown in the right panel of Fig.~\ref{fig4}(b), the decay rate progressively concentrates at the right intra-period time boundary, confirming the temporal NHSE with reversed chirality.

Despite their superficial similarities, the fundamental distinction between the temporal NHSE and its spatial counterpart lies in coupling directionality and the nature of the effective boundary. First, while spatial propagation is naturally bidirectional, temporal evolution strictly obeys causality. Although geometric-phase engineering allows leftward coupling ($i-1 \leftarrow i$) to dominate in the temporal NHSE, causality prohibits information from traveling backward in time. Governed by the discrete time-translation symmetry of the Floquet-Wannier basis, the system forms a temporal superlattice that repeats with period $T_0$, hosting $s$ sublattices per supercell [Fig.~\ref{fig4}(c)]. Because the characteristic hopping timescale satisfies $\hbar/\lvert\mathcal{J}_{\text{eff}}\rvert, \hbar/\lvert\Omega_{\text{eff}}^\prime\rvert \gg T_0$, this dominant leftward coupling spans across multiple supercell periods. Consequently, site $i-1$ responds to site $i$ with an intra-cell temporal offset of $-T_0/s$ but a net delay of $N T_0 - T_0/s$ ($N \ge 1$), guaranteeing that information flow always points strictly toward the future.

Second, unlike a spatial boundary that acts as a hard wall where non-reciprocal drift forces probability condensation, the temporal boundary here cannot obstruct the forward flow of time. Originating from the projection of the spatiotemporal boundary engineered in the FPS, it repeats periodically in each supercell across stroboscopic time windows $\tau \in [0,T_0)$ [Fig.~\ref{fig4}(a)], analogous to spatiotemporally modulated photonic crystals~\cite{5hf5-pg3t}. Specifically, it acts as a sharp energy step that rises periodically at position $\theta_0$ whenever $t = (\theta_0 - \Theta_e)/\Omega + N T_0$. This mechanism fundamentally differs from temporal interfaces created in recent observations of time reflection via rapid global parameter modulations~\cite{moussa2023observation, dong2024quantum}, which conserve momentum while breaking energy conservation. 
Instead of inducing time reflection, our boundary redirects the probability density toward regions ($\theta \neq \theta_0$) free of energy steps, allowing the wavepacket to bypass the barrier, enter the subsequent temporal supercell, and undergo skin accumulation iteratively.

To verify the dynamical control over the effective boundary locations across spatial and temporal domains and to assess the robustness of the temporal funneling, we perform numerical simulations under varying phase offsets $\Theta_e$ and random driving-frequency fluctuations in $F_{a,b}(t)$ (see Supplemental Materials~\cite{supplementarymaterial}). While the spatial NHSE is inherently robust against defects and disorder~\cite{weidemann2020topological, longhi2015robust, zheng2024dynamic}, we reveal that temporal funneling exhibits comparable robustness, corroborating the underlying nontrivial temporal topology.
Finally, we confirm the experimental feasibility using ultracold $^{87}\text{Rb}$ atoms in a ring trap of radius $r = 100\,\mu\text{m}$. For rotation period $T_0 = 0.54\,\text{ms}$, shaking period $T_s = 0.22\,\text{s}$, Rabi frequency $\Omega_R = 1.86\,\text{Hz}$, dissipation rate $\gamma = 0.46\,\text{Hz}$, and a required coherence time on the order of seconds, all parameters fall well within current experimental capabilities.

\emph{Conclusion}---This work extends the NHSE from conventional spatial dimensions to spatiotemporal dimensions in a Floquet-engineered atomic condensate. The spatial projection provides a new approach to boundary-free non-Hermitian quantum matter manipulation, while the temporal projection introduces causality-constrained time-domain nonreciprocity and periodic temporal funneling, opening an unexplored avenue for non-Hermitian topological physics. 
Compared with commonly used synthetic photonic mesh networks and photonic time crystals, the atomic condensate platform inherently supports quantum many-body interactions, offers versatile dynamic control over non-Hermitian dissipation and Floquet drive profiles, and avoids optical dispersion constraints and ultrafast parameter instabilities. It not only opens a route to high-resolution observation of genuine temporal skin effects but also lays the foundation for exploring non-Hermitian topological phenomena and quantum state engineering across coupled spatiotemporal dimensions.

\begin{acknowledgments}
\emph{Acknowledgements}---This work was supported by the
Quantum Science and Technology–National Science and
Technology Major Project (No. 2021ZD0303200); the
National Natural Science Foundation of China
(No. 12374328, No. 11974116, and No. 12234014); the
Science and Technology Innovation Plan of Shanghai
Science and Technology Commission (No. 24LZ1400600);
the Shanghai Municipal Science and Technology
Major Project (No. 2019SHZDZX01); the National Key
Research and Development Program of China
(No. 2016YFA0302001); the Fundamental Research Funds
for the Central Universities; the Chinese National Youth
Talent Support Program, and the Shanghai Talent program.
\end{acknowledgments}

\nocite{*}
\bibliography{Ref}

 \clearpage
\widetext
\begin{center}
    \textbf{\large Supplemental Materials}
\end{center}

\setcounter{equation}{0}
\setcounter{figure}{0}
\setcounter{table}{0}
\setcounter{page}{1}
\makeatletter
\renewcommand{\theequation}{S\arabic{equation}}
\renewcommand{\thefigure}{S\arabic{figure}}
\renewcommand{\bibnumfmt}[1]{[S#1]}
\renewcommand{\citenumfont}[1]{S#1}


\section{Effective lattice Hamiltonian in the Floquet phase space}

We employ a semiclassical Floquet-phase-space (FPS) approach, in which the time-dependent classical Hamiltonian is first subjected to a canonical transformation in the extended phase space, followed by quantization of the resulting slow degrees of freedom. This approach, widely used in atomic time-crystal studies, provides direct access to the underlying classical phase-space structure while retaining the relevant quantum dynamics, offering a complementary perspective to fully quantum Floquet theory.

We first consider the classical center-of-mass dynamics in the laboratory frame. In the absence of Rabi coupling and dissipation, the classical Hamiltonian for the two internal states $a$ and $b$ is
\begin{equation}
H_{a,b}(\theta,p,t)=
\frac{p^2}{2Mr^2}+\mathcal{V}_{a,b}(\theta,t).
\end{equation}
The corresponding unperturbed dynamics of an ultracold atom with angular momentum $p_0=Mrv$ is characterized by $\frac{dp}{dt}=0$, $\frac{d\theta}{dt}=\Omega=\frac{v}{r}$. The weak time-dependent potentials $\mathcal{V}_{a,b}(\theta,t)$ therefore act as perturbations to the uniform rotational motion. As detailed below, when the driving frequency $\omega$ is a harmonic of $\Omega$, the near-resonant Fourier components can accumulate coherently and modify the atomic dynamics, whereas rapidly oscillating off-resonant components average out.

In this study, we engineer the time-dependent driving functions entering the perturbations, $\mathcal{V}_{a,b}(\theta,t)=V(\theta)F_{a,b}(t)$, as
\begin{eqnarray}
    F_a(t)&=&\cos(\omega t+\phi(t)+\Theta_a)+f_e(t),\\
    F_b(t)&=&\cos(\omega t+\phi(t)+\Theta_b)+f_e(t).
\end{eqnarray}
Here, the phase-shaking frequency satisfies $\omega_s \ll \Omega \ll \omega$, and $f_e(t)$ is a generalized boundary-control function. Under a secular approximation, the harmonic cosine terms generate a dynamic shaking lattice, whereas a tailored $f_e$ establishes the effective edges in the Floquet phase space (FPS).

To illustrate this, we expand $\exp[\pm i\phi(t)]$ into a temporal Fourier series with shaking frequency $\omega_s$. For $\omega = m_0 \omega_s$, the cosine component in $F_a(t)$ becomes
\begin{equation}
\cos[\omega t+\phi(t)+\Theta_a] = \frac{e^{i\Theta_a}}{2} \sum_m c_m e^{i(m+m_0)\omega_s t} + \frac{e^{-i\Theta_a}}{2} \sum_{m'} c'_{m'} e^{i(m'-m_0)\omega_s t},
\end{equation}
where $c_m = \frac{1}{T_s}\int_0^{T_s} dt\, e^{i\phi(t)} e^{-im \omega_s t}$ and $c'_{m'} = \frac{1}{T_s}\int_0^{T_s} dt\, e^{-i\phi(t)} e^{-im' \omega_s t}$. Combining this with the spatial Fourier series $V(\theta) = \sum_n V_n e^{in\theta}$ yields terms with spatiotemporal phases $in\theta + i(m \pm m_0)\omega_s t \pm i\Theta_a$. Along the unperturbed trajectory $d\theta/dt \approx \Omega = \omega/s = m_0\omega_s/s$, the slow-varying (near-resonant) terms satisfy $n \Omega + (m \pm m_0)\omega_s \approx 0$. Since $\omega_s \ll \Omega \ll \omega$, the low-frequency components ($m, m' \ll m_0$) dominate, while higher-order Fourier coefficients decay rapidly. Thus, the resonance condition simplifies to $n \approx \mp s$, isolating the primary resonant harmonic $V_s$. Performing the summation under the secular approximation, we obtain
\begin{equation}
V(\theta)\cos[\omega t+\phi(t)+\Theta_a] \approx  V_s \cos[s(\theta-\Omega t)-\phi(t)-\Theta_a].
\end{equation}
Applying the same secular approximation to $F_b(t)$ yields an identical result, with $\Theta_a$ replaced by $\Theta_b$.

Similarly, for the boundary-engineering term $V(\theta)f_e(t)$ in $\mathcal{V}_{a,b}(\theta,t)$, expanding both $V(\theta) = \sum_n V_n e^{in\theta}$ and $f_e(t) = \sum_m F_m e^{im(\Omega t + \Theta_e)}$ with an adjustable phase offset $\Theta_e$ yields
\begin{equation}
V(\theta)f_e(t) = \sum_{n,m} V_n F_m e^{in\theta} e^{im(\Omega t + \Theta_e)}.
\end{equation}
Isolating the resonant terms ($m = -n$) along the unperturbed trajectory via the same secular approximation gives
\begin{equation}
V(\theta)f_e(t) \approx \sum_n V_n F_{-n} e^{in(\theta - \Omega t - \Theta_e)}.
\end{equation}
To engineer effective boundary edges, setting $V_n F_{-n} \propto e^{-n^2 / 2a^2}$ constructs a localized Gaussian barrier:
\begin{equation}
V(\theta)f_e(t) \approx V_e \exp\left[-\frac{a^2}{2}(\theta - \Omega t - \Theta_e)^2\right].
\end{equation}
Other boundary profiles can be flexibly customized by tailoring the coefficient relation $V_n F_{-n}$.

To isolate the perturbation-induced dynamics, we perform a canonical transformation to the rotating frame,
\begin{equation}
\bar{\theta}=\theta-\Omega t,
\qquad
\bar p=p,
\qquad
P_t\rightarrow P_t-\Omega p.
\end{equation}
The extended Hamiltonian therefore becomes
\begin{equation}
H_{a,b}^{\prime}=\frac{p^2}{2Mr^2}-\Omega p+P_t
+\mathcal{V}_{a,b}(\bar{\theta},t),
\end{equation}
where $P_t$ is the canonical momentum conjugate to $t$. The effective potentials are
\begin{eqnarray}
\mathcal{V}_{a,b}(\bar{\theta},t)
&=&
V_s\cos[s\bar{\theta}-\phi(t)-\Theta_{a,b}]
+V_e(\bar{\theta}),\\
V_e(\bar{\theta})
&=&
V_e\exp\left[-\frac{a^2}{2}(\bar{\theta}-\Theta_e)^2\right].
\end{eqnarray}

Since the perturbation is weak relative to the rotational kinetic energy, we expand $p$ around the unperturbed value $p_0 = M r^2 \Omega$. Expanding the kinetic term to second order in $P = p - p_0$, the linear cross-term $\frac{p_0}{Mr^2}P = \Omega P$ cancels the $-\Omega p$ shift (up to an omitted constant energy offset), simplifying the Hamiltonian to
\begin{equation}H_{a,b}^{\prime} \approx \frac{P^2}{2M r^2} + \mathcal{V}_{a,b}(\bar\theta,t) + P_t.\end{equation}
Here, the rotating-frame coordinate $\bar\theta=\theta-\Omega t$ mixes the spatial coordinate and time linearly, placing them on an equal footing in the extended phase space (FPS), while $(P,\bar\theta)$ form the corresponding slow canonical pair.

\begin{figure}[t]
	\centering
	\includegraphics[width=0.8\linewidth, height=0.47\linewidth]{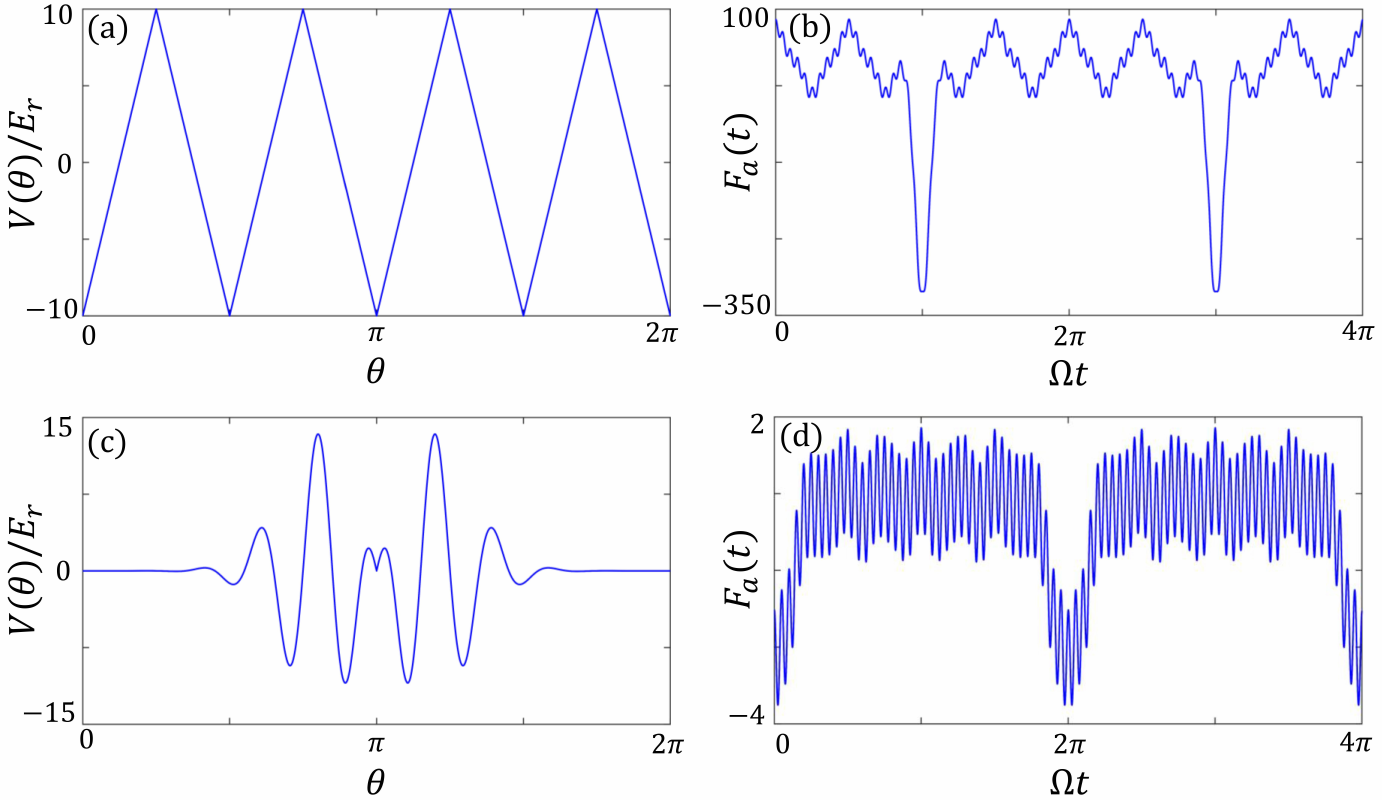}
	\caption{(a),(c) Triangular periodic and nonperiodic arbitrary perturbation potentials $V(\theta)$, alongside (b),(d) their respective temporal driving functions $F_a(t)$ ($\Theta_e=0$). Remarkably, despite distinct $V(\theta)$, both driving functions construct an identical effective Gaussian barrier in the FPS.}
\end{figure}

For $V_s \ll V_e \ll E_{\text{rot}}=\frac{p_0^2}{2Mr^2}$ and a width much smaller than $2\pi$, the Gaussian potential $V_e(\bar{\theta})$ acts as an effective localized boundary at $\bar{\theta}=\Theta_e$ in the FPS, preserving the validity of the perturbative expansion. As illustrated in FIG.S1 for two distinct spatial potentials $V(\theta)$ and their driving functions $F_a(t)$ ($F_b(t)$ being analogous), neither $V(\theta)$ breaks the physical periodic boundary condition (PBC) along the spatial ring $\theta$. Instead, the tailored driving $F_a(t)$ dynamically constructs this Gaussian barrier in the FPS, converting the spatial PBC in $\theta$ into a generalized or open boundary condition in $\bar{\theta}$.

Quantizing the slow canonical pair $(P,\bar\theta)$ while treating $t$ as the physical time parameter, and incorporating the Rabi coupling $\Omega_R$ and dissipation $\gamma$, gives the effective quantum Hamiltonian
\begin{equation}
\mathcal{H}_0(t)=
\begin{pmatrix}
\frac{\hat P^2}{2Mr^2}+\mathcal{V}_a(\bar\theta,t) & \hbar\Omega_R \\
\hbar\Omega_R &
\frac{\hat P^2}{2Mr^2}+\mathcal{V}_b(\bar\theta,t)-i\hbar\gamma
\end{pmatrix}.
\end{equation}
Here, the $P_t$ term of the extended-space Hamiltonian has been absorbed into the time-evolution operator, $P_t\rightarrow-i\hbar\partial_t$, so that the dynamics are governed by $i\hbar\partial_t|\Psi\rangle=\mathcal{H}_0(t)|\Psi\rangle$.

To eliminate the lattice shaking, we perform a unitary transformation $U_1 = \exp[-i Mr^2\dot{\phi}(t)\bar{\theta}/s\hbar] \exp[-i\phi(t)\hat P/s\hbar]$. Neglecting a uniform energy shift, the transformed Hamiltonian reads
\begin{equation}
\mathcal{H}_1=
\begin{pmatrix}
\frac{\hat P^2}{2Mr^2}+\mathcal{V}_a(\bar{\theta})+\mathcal{F}(t)\bar{\theta} & \hbar\Omega_R \\
\hbar\Omega_R & \frac{\hat P^2}{2Mr^2}+ \mathcal{V}_b(\bar{\theta})+\mathcal{F}(t)\bar{\theta}-i\hbar\gamma
\end{pmatrix},
\end{equation}
where $\mathcal{V}_{a,b}(\bar\theta) = V_s\cos(s\bar{\theta}-\Theta_{a,b}) + V_e(\bar{\theta})$, and $\mathcal{F}(t) = \frac{Mr^2}{s}\frac{d^2\phi(t)}{dt^2}$ is an inertial force. Since the shaking amplitude is sub-lattice-period, its dynamic effect on $V_e(\bar{\theta})$ is negligible.

Under a deep lattice depth $V_s$, setting $\Theta_b-\Theta_a=\pi$ to construct the zigzag ladder and neglecting higher bands and expanding atomic field operators in the localized Floquet-Wannier basis $\mathcal{W}_{i}(\bar{\theta})$ (having a spatial period $d=2\pi/s$, corresponding to a temporal period $T_0/s$ in the laboratory frame) yields the tight-binding Hamiltonian
\begin{align}
\mathcal{H}_1 &\approx -\mathcal{J}\sum_{i} (a_i^\dagger a_{i+1} + b_i^\dagger b_{i+1} + \text{H.c.}) + \hbar\Omega_R^\prime\sum_{\langle i,j\rangle} [a_i^\dagger(b_j + b_{j+1}) + \text{H.c.}] \nonumber \\
&\quad + \mathcal{F}(t)\sum_{i} r_i (n_i^a + n_i^b) - i\hbar\gamma\sum_{i} n_i^b + \mathcal{H}_e,
\end{align}
where $a_i (b_i)$ are annihilation operators, $\mathcal{J}$ is the intra-leg nearest-neighbor tunneling rate, $\Omega_R^\prime = \alpha\Omega_R$ with overlap integral $\alpha = \int d\bar\theta \mathcal{W}_j^*(\bar\theta)\mathcal{W}_i(\bar\theta)$ is the inter-leg coupling of the zigzag ladder, and $r_i$ denotes Wannier center positions. 
The boundary Hamiltonian $\mathcal{H}_e \approx \mathcal{J}_e (a_{i_e}^\dagger a_{i_e+1} + b_{i_e}^\dagger b_{i_e+1} + \text{H.c.})$ captures the localized dynamics, modified specifically across the boundary adjacent sites $i_e$ and $i_e+1$ near $\bar\theta=\Theta_e$, where $\mathcal{J}_e = \int d\bar{\theta}\mathcal{W}_{i_e}^*(\bar{\theta})[\frac{P^2}{2Mr^2}+V_e(\bar{\theta})]\mathcal{W}_{i_e+1}(\bar{\theta})$.

To map the inertial drive into time-dependent Peierls phases, we apply a second transformation $U_2 = \exp[\frac{i}{\hbar}\sum_i \chi_i(t)(n_i^a + n_i^b)]$ with $\chi_i(t) = -r_i \int_{t_0}^t dt' \mathcal{F}(t') + \frac{r_i}{T_s}\int_0^{T_s} dt \int_{t_0}^t dt' \mathcal{F}(t')$. This maps the tunneling terms to
\begin{align}
\mathcal{H}_2 &= -\mathcal{J}\sum_{i} \left( e^{i\varphi_{i,i+1}(t)} a_i^\dagger a_{i+1} + e^{i\varphi_{i,i+1}(t)}b_i^\dagger b_{i+1} + \text{H.c.} \right) \nonumber \\
&\quad + \hbar\Omega_R^\prime\sum_{\langle i,j \rangle} \left[ a_i^\dagger (e^{i\varphi^{\prime}_{i,j}(t)}b_j + e^{i\varphi^{\prime}_{i,j+1}(t)}b_{j+1}) + \text{H.c.} \right] - i\hbar\gamma\sum_{i}n_i^b + \mathcal{H}_e,
\end{align}
where $\varphi_{i,i+1}(t) = [\chi_{i+1}(t)-\chi_i(t)]/\hbar$ and $\varphi'_{i,j}(t) = [\chi_j(t)-\chi_i(t)]/\hbar$. Because boundary tunneling rates in $\mathcal{H}_e$ are much weaker than $\mathcal{J}$ and $\Omega_R^\prime$, $\mathcal{H}_e$ remains invariant under $U_2$.

\begin{figure}[t]
	\centering
	\includegraphics[width=0.5\linewidth, height=0.4\linewidth]{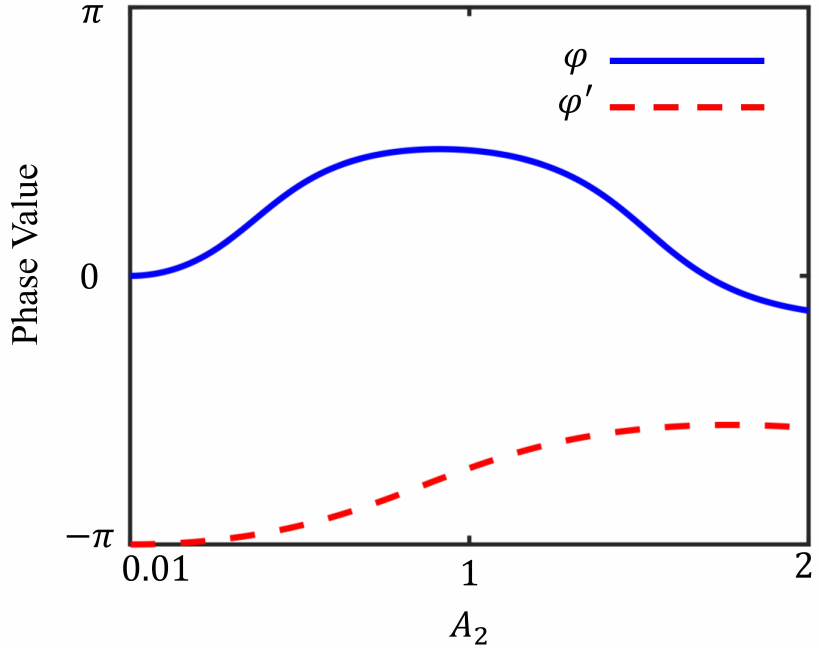}
	\caption{Synthetic Peierls phases $\varphi$ and $\varphi'$ as a function of shaking amplitude $A_2$ at fixed $A_1 = 0.95$.}
\end{figure}

For a bi-harmonic phase drive 
\begin{equation}
    \phi(t) = A_1\sin(2\omega_s t) + A_2\cos(\omega_s t)
\end{equation}
with frequency $\omega_s$ far exceeding the lowest two band widths $W_{1,2}$ yet well below the excited-band gap ($\hbar\Delta_{\text{gap}} \gg \hbar\omega_s \gg W_{1,2}$), high-frequency Floquet averaging yields the static effective Hamiltonian
\begin{equation}
\mathcal{H}_2 \approx -|\mathcal{J}_{\text{eff}}|\sum_{i} (e^{i\varphi}a_i^\dagger a_{i+1} + e^{i\varphi}b_i^\dagger b_{i+1} + \text{H.c.}) + \hbar|\Omega_{\text{eff}}^\prime|\sum_{\langle i,j \rangle}[a_i^\dagger(e^{-i\varphi^{\prime}}b_j + e^{i\varphi^{\prime}}b_{j+1}) + \text{H.c.}] - i\hbar\gamma\sum_i n_i^b + \mathcal{H}_e.
\end{equation}
The renormalized parameters are given by generalized Bessel function ($J_m$) expansions:
\begin{align}
\frac{\mathcal{J}_{\text{eff}}}{\mathcal{J}} &= \sum_{m=-\infty}^{\infty} (-i)^m J_m\left(\frac{2M r^2 \omega_s A_1 d}{s\hbar}\right) J_{2m}\left(\frac{M r^2\omega_s A_2 d}{s\hbar}\right) = |\mathcal{J}_{\text{eff}}/\mathcal{J}| e^{i\varphi}, \\
\frac{\Omega_{\text{eff}}^\prime}{\Omega_R^\prime} &= \sum_{m=-\infty}^{\infty} (-i)^m J_m\left(\frac{M r^2 \omega_s A_1 d}{s\hbar}\right) J_{2m}\left(\frac{M r^2 \omega_s A_2 d}{2s \hbar\omega_s}\right) = |\Omega_{\text{eff}}^\prime/\Omega_R^\prime| e^{i\varphi'},
\end{align}
where $\varphi = \text{Arg}(\mathcal{J}_{\text{eff}}/\mathcal{J})$ and $\varphi' = \text{Arg}(\Omega_{\text{eff}}^\prime/\Omega_R^\prime)$ represent controllable synthetic gauge phases. As shown in FIG. S2, these synthetic phases can be continuously tuned by modulating $A_2$ at fixed $A_1 = 0.95$.

\section{Adiabatic Elimination Analysis of Nonreciprocal Coupling}
To clarify the origin of nonreciprocity induced by atomic dissipation and the synthetic geometric phase, we adiabatically eliminate the fast-decaying internal state $b$ under the strong-dissipation regime $\hbar\gamma \gg \hbar \vert{}\Omega_{\text{eff}}'\vert{}, \vert{}\mathcal{J}_{\text{eff}}\vert{}$. 

Assuming periodic boundary conditions in the FPS, the tight-binding Hamiltonian $\mathcal{H}_2$ in quasimomentum $k$-space reads
\begin{equation}
\mathcal{H}_k = \sum_{k} \left[ \epsilon_k a_k^\dagger a_k + (\epsilon_k - i\hbar\gamma) b_k^\dagger b_k + \hbar \tilde{\Omega}_k (a_k^\dagger b_k + \text{H.c.}) \right],
\end{equation}
where $\epsilon_k \equiv \epsilon_k^{a,b} = -2\vert{}\mathcal{J}_{\text{eff}}\vert{}\cos(kd+\varphi)$ represents the bare band dispersion, and $\tilde{\Omega}_k = 2\vert{}\Omega_{\text{eff}}'\vert{}\cos(kd/2+\varphi')$ is the $k$-dependent inter-state coupling.

The corresponding Heisenberg equations for $a_k$ and $b_k$ are cleanly expressed as
\begin{eqnarray}
i\hbar\frac{\partial}{\partial t} a_k &=& \epsilon_k a_k + \hbar\tilde{\Omega}_k b_k, \\
i\hbar\frac{\partial}{\partial t} b_k &=& (\epsilon_k - i\hbar\gamma) b_k + \hbar\tilde{\Omega}_k a_k.
\end{eqnarray}
Under $\hbar\gamma \gg \hbar \vert{}\Omega_{\text{eff}}'\vert{}, \vert{}\mathcal{J}_{\text{eff}}\vert{}$, setting $\partial b_k / \partial t \approx 0$ allows adiabatically eliminating $b_k$, yielding $b_k = \frac{\hbar\tilde{\Omega}_k}{i\hbar\gamma - \epsilon_k} a_k$. Substituting this back gives
\begin{equation}
i\hbar\frac{\partial}{\partial t} a_k = \left( \epsilon_k + \frac{\hbar^2 \tilde{\Omega}_k^2}{i\hbar\gamma - \epsilon_k} \right) a_k.
\end{equation}
This dictates an effective single-band Hamiltonian $\mathcal{H}_k^{\prime} = \sum_k E_k a_k^\dagger a_k$. Expanding the energy $E_k$ in powers of $\epsilon_k / (\hbar\gamma) \ll 1$ leads to
\begin{equation}
E_k = \epsilon_k + \frac{\hbar^2 \tilde{\Omega}_k^2}{i\hbar\gamma} \left( 1 - \frac{\epsilon_k}{i\hbar\gamma} \right)^{-1} \approx \epsilon_k + \frac{\hbar \tilde{\Omega}_k^2}{i\gamma} - \frac{\tilde{\Omega}_k^2 \epsilon_k}{\gamma^2}.
\end{equation}

To manifest the synthetic geometric phase $\Phi = 2\varphi' - \varphi$, we use the identity $\tilde{\Omega}_k^2 = 2\vert{}\Omega_{\text{eff}}'\vert{}^2 [\cos(kd+\varphi+\Phi)+1]$. Substituting the explicit forms of $\epsilon_k$ and $\tilde{\Omega}_k^2$ into $E_k$, and omitting an overall energy constant, the approximated Hamiltonian expands to
\begin{gather}
    \begin{aligned}
        \mathcal{H}_k^{\prime}\approx &\sum_k (-2|\mathcal{J}_{\text{eff}}|+\frac{2\hbar\vert{}\Omega_{\text{eff}}'\vert{}^2\cos\Phi}{i\gamma}+\frac{4\vert{}\Omega_{\text{eff}}'\vert{}^2 |\mathcal{J}_{\text{eff}}|}{\gamma^2})\cos(kd+\varphi)a_k^\dagger a_k-\sum_k \frac{2\hbar\vert{}\Omega_{\text{eff}}'\vert{}^2}{i\gamma}\sin\Phi\sin(kd+\varphi) a_k^\dagger a_k\\
        &+\sum_k \frac{4\vert{}\Omega_{\text{eff}}'\vert{}^2 |\mathcal{J}_{\text{eff}}|}{\gamma^2}[\cos^2(kd+\varphi)\cos\Phi-\cos(kd+\varphi)\sin(kd+\varphi)\sin\Phi]a_k^\dagger a_k,
    \end{aligned}
\end{gather}

Transforming back to the real-space lattice configuration, we obtain
\begin{gather}
    \begin{aligned}
        \mathcal{H}_{\textbf{eff}}\approx &\sum_i (-|\mathcal{J}_{\text{eff}}|+\frac{\hbar\vert{}\Omega_{\text{eff}}'\vert{}^2\cos\Phi}{i\gamma}+\frac{\hbar\vert{}\Omega_{\text{eff}}'\vert{}^2\sin\Phi}{\gamma}+\frac{2\vert{}\Omega_{\text{eff}}'\vert{}^2 |\mathcal{J}_{\text{eff}}|}{\gamma^2})e^{i\varphi}a_i^\dagger a_{i+1}\\
        &+\sum_i(-|\mathcal{J}_{\text{eff}}|+\frac{\hbar\vert{}\Omega_{\text{eff}}'\vert{}^2\cos\Phi}{i\gamma}-\frac{\hbar\vert{}\Omega_{\text{eff}}'\vert{}^2\sin\Phi}{\gamma}+\frac{2\vert{}\Omega_{\text{eff}}'\vert{}^2 |\mathcal{J}_{\text{eff}}|}{\gamma^2})e^{-i\varphi}a_{i+1}^\dagger a_{i}\\
        &+\sum_i (\frac{\vert{}\Omega_{\text{eff}}'\vert{}^2 |\mathcal{J}_{\text{eff}}|}{\gamma^2}\cos\Phi-\frac{\vert{}\Omega_{\text{eff}}'\vert{}^2 |\mathcal{J}_{\text{eff}}|}{i\gamma^2}\sin\Phi)e^{2i\varphi} a_i^\dagger a_{i+2}\\
        &+\sum_i(\frac{\vert{}\Omega_{\text{eff}}'\vert{}^2 |\mathcal{J}_{\text{eff}}|}{\gamma^2}\cos\Phi+\frac{\vert{}\Omega_{\text{eff}}'\vert{}^2 |\mathcal{J}_{\text{eff}}|}{i\gamma^2}\sin\Phi)e^{-2i\varphi} a_{i+2}^\dagger a_{i}.
    \end{aligned}
\end{gather}

The interplay between dissipation $\gamma$ and a non-zero geometric phase $\Phi \neq 0$ manifests directly as explicit directional nonreciprocity ($t_R \neq t_L^*$), where the forward and backward complex hopping amplitudes are given by
\begin{align}
t_R &= \left( -|\mathcal{J}_{\text{eff}}| + \frac{\hbar|\Omega_{\text{eff}}'|^2\cos\Phi}{i\gamma} - \frac{\hbar|\Omega_{\text{eff}}'|^2\sin\Phi}{\gamma} + \frac{2|\Omega_{\text{eff}}'|^2 |\mathcal{J}_{\text{eff}}|}{\gamma^2} \right) e^{-i\varphi}, \\
t_L &= \left( -|\mathcal{J}_{\text{eff}}| + \frac{\hbar|\Omega_{\text{eff}}'|^2\cos\Phi}{i\gamma} + \frac{\hbar|\Omega_{\text{eff}}'|^2\sin\Phi}{\gamma} + \frac{2|\Omega_{\text{eff}}'|^2 |\mathcal{J}_{\text{eff}}|}{\gamma^2} \right) e^{i\varphi}.
\end{align}

\section{ANALYSES OF CONTROLLABILITY AND ROBUSTNESS }

\begin{figure}[t]
	\centering
	\includegraphics[width=0.7\linewidth, height=0.26\linewidth]{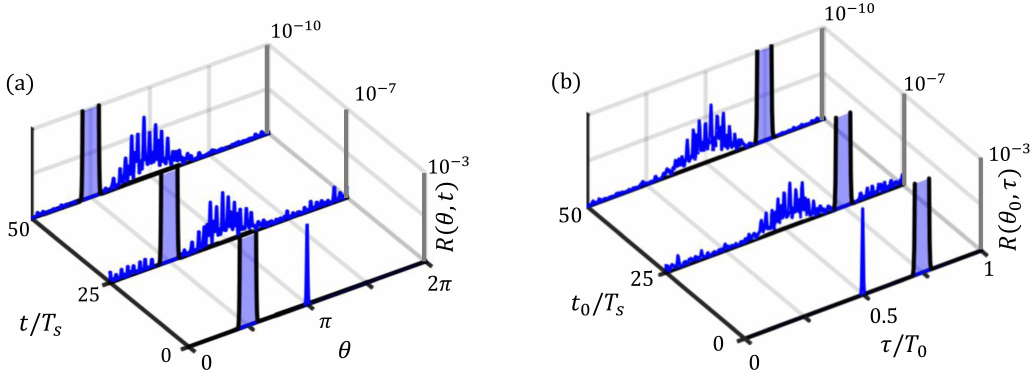}
	\caption{Spatiotemporal decay rates (a) $R(\theta,t)$ and (b) $R(\theta_0,\tau)$ under a shifted driving phase $\Theta_e = -3\pi/2$ in the presence of temporal frequency disorder $\delta\omega$. All other parameters are identical to those in Fig.~3(b) of the main text.}
\end{figure}

Since the spatial location of the effective boundary $V_e(\bar{\theta})$ is continuously tunable via the driving phase $\Theta_e$, we verify this boundary reconfigurability in the laboratory frame by conducting numerical simulations identical to those in Fig.~\ref{fig4}(b) of the main text. Specifically, we evaluate the time-dependent Schrödinger equation
\begin{equation}
i\hbar\frac{\partial}{\partial t}\psi(\theta,t) = \mathcal{H}_0(t)\psi(\theta,t)
\end{equation}
starting from the same initial state $\psi(\theta,0) = [\psi_a(\theta,0), \psi_b(\theta,0)]^T$, where $\psi_a(\theta,0)$ is a Gaussian wavepacket centered at $\theta = \pi$ with an average angular momentum $p_0$, and $\psi_b(\theta,0) = 0$.

To demonstrate phase-controlled boundary relocation, the phase parameter $\Theta_e$ in the driving modulation
\begin{equation}
f_e(t) = \sum_m F_m \exp(im\Omega t + im\Theta_e)
\end{equation}
is shifted from $\Theta_e = 0$ in main text to $\Theta_e = -3\pi/2$, resulting in the effective boundary center located at $\bar\theta+3\pi/2=0$ in the FPS. Consequently, the spatial edge shifts from $\theta = 0$ to $\theta = \pi/2$, accompanied by a corresponding temporal boundary shift from $\tau = T_0$ to $\tau = 0.75 T_0$ for $\theta_0=0$.

Furthermore, to test the robustness of the spatiotemporal non-Hermitian skin effect (NHSE) against environmental fluctuations, we consider disorder in the temporal domain. Since the spatial potential $V(\theta)$ can already accommodate arbitrary non-lattice spatial perturbations, we focus on temporal noise within the driving functions $F_{a,b}(t)$ to evaluate temporal robustness. Concrete numerical simulations are performed by introducing stochastic frequency fluctuations into the driving term
\begin{equation}
\cos[\omega t + \phi(t)] \longrightarrow \cos[(\omega + \delta\omega) t + \phi(t)],
\end{equation}
where $\delta\omega$ is a zero-mean random variable with a characteristic noise amplitude of $\vert{}\mathcal{J}_{\text{eff}}\vert{} / (5\hbar)$.

The corresponding simulation results are depicted in FIGs.~S3(a) and S3(b). The persistent directional drift along with asymmetric decay dynamics confirm the preservation of spatiotemporal funneling under temporal noise and boundary relocation. These findings unambiguously confirm the flexible controllability of skin boundary positions and the strong physical robustness of the spatiotemporal NHSE.

\end{document}